\documentclass[english,aip,apl,reprint,superscriptaddress,showpacs]{revtex4-1}
\usepackage{mathptmx}
\usepackage[scaled=0.9]{helvet}

\usepackage[T1]{fontenc}
\usepackage[utf8]{inputenc}
\usepackage{color}
\usepackage{babel}

\usepackage{prettyref}
\usepackage{textcomp}
\usepackage{amstext}
\usepackage{graphicx}
\usepackage{amssymb}
\usepackage[unicode=true, 
 bookmarks=false,
 breaklinks=false,pdfborder={0 0 0},backref=false,colorlinks=true]
 {hyperref}
\hypersetup{pdftitle={Compensation-dependence of magnetic and electrical properties in GaMnP},
 pdfauthor={T. E. Winkler, P. R. Stone, Tian Li, K. M. Yu, A. Bonanni, O. D. Dubon},
 urlcolor=blue,citecolor=blue,linkcolor=blue}

\makeatletter
\@ifundefined{textcolor}{}
{%
 \definecolor{BLACK}{gray}{0}
 \definecolor{WHITE}{gray}{1}
 \definecolor{RED}{rgb}{1,0,0}
 \definecolor{GREEN}{rgb}{0,1,0}
 \definecolor{BLUE}{rgb}{0,0,1}
 \definecolor{CYAN}{cmyk}{1,0,0,0}
 \definecolor{MAGENTA}{cmyk}{0,1,0,0}
 \definecolor{YELLOW}{cmyk}{0,0,1,0}
 }

\usepackage{microtype}
\newrefformat{fig}{Fig.~\ref{#1}}
\AtBeginDocument{}

\makeatother

\begin{document}

\title{Compensation-dependence of magnetic and electrical properties in
Ga$_{\text{1-x}}$Mn$_{\text{x}}$P}

\author{T. E. Winkler}

\affiliation{Department of Materials Science and Engineering, University of California,
Berkeley, CA 94720, USA}

\affiliation{Materials Sciences Division, Lawrence Berkeley National Laboratory,
Berkeley, CA 94720, USA}

\affiliation{Institut für Halbleiter- und Festkörperphysik, Johannes Kepler University,
Altenbergerstraße 69, A-4040 Linz, Austria}

\author{P. R. Stone}

\affiliation{Department of Materials Science and Engineering, University of California,
Berkeley, CA 94720, USA}

\affiliation{Materials Sciences Division, Lawrence Berkeley National Laboratory,
Berkeley, CA 94720, USA}

\author{Tian Li}

\affiliation{Institut für Halbleiter- und Festkörperphysik, Johannes Kepler University,
Altenbergerstraße 69, A-4040 Linz, Austria}

\author{K. M. Yu}

\affiliation{Department of Materials Science and Engineering, University of California,
Berkeley, CA 94720, USA}

\affiliation{Materials Sciences Division, Lawrence Berkeley National Laboratory,
Berkeley, CA 94720, USA}

\author{A. Bonanni}

\email{alberta.bonanni@jku.at}

\affiliation{Institut für Halbleiter- und Festkörperphysik, Johannes Kepler University,
Altenbergerstraße 69, A-4040 Linz, Austria}

\author{O. D. Dubon}

\email{oddubon@berkeley.edu}

\affiliation{Department of Materials Science and Engineering, University of California,
Berkeley, CA 94720, USA}

\affiliation{Materials Sciences Division, Lawrence Berkeley National Laboratory,
Berkeley, CA 94720, USA}

\date{October 6, 2010}
\begin{abstract}
We demonstrate the control of the hole concentration in Ga$_{1-x}$Mn$_{x}$P
over a wide range by introducing compensating vacancies. The resulting
evolution of the Curie temperature from $51$~$\mathrm{K}$ to $7.5$~$\mathrm{K}$
is remarkably similar to that observed in Ga$_{1-x}$Mn$_{x}$As despite
the dramatically different character of hole transport between the
two material systems. The highly localized nature of holes in Ga$_{1-x}$Mn$_{x}$P
is reflected in the accompanying increase in resistivity by many orders
of magnitude. Based on variable-temperature resistivity data we present
a general picture for hole conduction in which variable-range hopping
is the dominant transport mechanism in the presence of compensation.
\end{abstract}


\maketitle
Dilute magnetic semiconductors (DMSs), where a few atomic percent
of magnetic ions are randomly substituted for a semiconductor host
species, represent a remarkable workbench for the study and demonstration
of spintronic functionalities.\citep{Dietl_08} They are not only
a means to an end but very exciting materials in their own right,
exhibiting many striking phenomena whose interpretation and modeling
are extremely challenging -- from the ferromagnetic exchange itself
to the large anomalous Hall effect.\citep{Jungwirth_RMP06} Much research
has focused on III-Mn-V systems,\citep{Jungwirth_RMP06,Matsukura_02,Burch_JMMM08,Sato_RMP10}
where Mn acts as the source of both magnetic moment and carriers that
mediate long-range ordering. While the behavior of Ga$_{1-x}$Mn$_{x}$As
is reasonably well understood at this point, the models developed
in the process fall short of describing some other DMSs.\citep{Jungwirth_RMP06}

Ga$_{1-x}$Mn$_{x}$P is a prime candidate for further study, due
to both its chemical similarity to Ga$_{1-x}$Mn$_{x}$As as well
as its low $0.36$~$\mathrm{\%}$ lattice mismatch with Si. Because
the Mn acceptor level lies approximately four times deeper within
the gap with respect to the valence band than in GaAs,\citep{Clerjaud_JPC85}
the holes are of a much more localized nature. Still, hole-mediated
ferromagnetism (FM) has been demonstrated conclusively in Ga$_{1-x}$Mn$_{x}$P
fabricated by ion implantation and pulsed-laser melting (II-PLM).\citep{Scarpulla_PRL05}
In the best samples to date FM signatures persist up to a Curie temperature
($T_{\mathrm{C}}$) of $65$~$\mathrm{K}$,\citep{Farshchi_SSC06}
which is $25$~$\mathrm{K}$ lower than for Ga$_{1-x}$Mn$_{x}$As
at the same $x=0.042$.\citep{Jungwirth_PRB05}

One of the hallmarks of carrier-mediated FM is the dependence of the
characteristic electrical, magnetic and optical properties on $x$
and carrier (\textsl{i.e.}, hole) concentration, $p$. A major line
of study pursued has thus been the behavior of Ga$_{1-x}$Mn$_{x}$P
over a range of $x$.\citep{Stone_APL06,Farshchi_SSC06} While these
samples implicitly exhibit different $p$ as well, this approach only
explores part of the available parameter space. Research into samples
with constant $x$ and varying $p$ has been comparatively limited,
focusing on anisotropy in S-codoped samples\citep{Stone_PRB08} and
on $T_{\mathrm{C}}$ in S- and Te-codoped samples.\citep{Scarpulla_JAP08a,Scarpulla_PRL05}

In this letter we present the first systematic study on the electrical
and magnetic effects of hole compensation in Ga$_{1-x}$Mn$_{x}$P.
We utilize the amphoteric nature of native defects \citep{Walukiewicz_APL89}
-- donor-like in Ga$_{1-x}$Mn$_{x}$P \citep{Walukiewicz_PRB88,Clerjaud_JPC85}
-- to investigate a very wide range of $p$ without significantly
changing $x$. A similar method has recently been applied to Ga$_{1-x}$Mn$_{x}$As,\citep{Mayer_PRB10}
and we find surprising similarities between the materials despite
the radically different degree of hole localization. Furthermore,
we present a picture for hole conduction by variable-range hopping
(VRH) in Ga$_{1-x}$Mn$_{x}$P.

The samples for this study were prepared by II-PLM.\citep{Scarpulla_APL03}
A GaP $\bigl(001\bigr)$ wafer -- doped n-type; $n\sim10^{16}\text{--}10^{17}$~$\mathrm{cm^{-3}}$
-- was implanted with $\mathrm{Mn^{+}}$ at an energy of $50$~$\mathrm{keV}$
and an angle of incidence of $7\text{\textdegree}$ to a dose of $2\times10^{16}$~$\mathrm{cm^{-2}}$.
Samples with approximate side lengths of $6$~$\mathrm{mm}$ were
cleaved along $\bigl\langle110\bigr\rangle$ directions and individually
irradiated with a single $\sim0.4$~$\mathrm{J\, cm^{-2}}$ KrF laser
pulse ($\lambda=248$~$\mathrm{nm}$, $\mathrm{FWHM}=18$~$\mathrm{ns}$),
homogenized to a spatial uniformity of $\pm5$~$\%$ by a crossed-cylindrical
lens homogenizer. They were subsequently subjected to $24$~$\mathrm{h}$
HCl etching to remove residual surface damage. These parameters have
been used previously to produce samples with $x\approx0.038$.\citep{Farshchi_SSC06}
For our samples, $x$ is defined as the peak substitutional manganese
($\mathrm{Mn_{Ga}}$) fraction -- occurring between $20$ and $30$~$\text{nm}$
below the surface -- as determined by a combination of secondary ion
mass spectrometry (SIMS) and ion beam analysis (IBA).\citep{Stone_10}
Compensating defects were then introduced into samples by consecutive
irradiations with $\mathrm{Ar^{+}}$ at an energy of $33$~$\mathrm{keV}$
and an angle of incidence of $7\text{\textdegree}$, which according
to simulations \citep{Ziegler_08} yield a vacancy depth profile similar
to the typical Mn distribution.

The characterization of several identically prepared Ga$_{1-x}$Mn$_{x}$P
samples was carried out by SQUID magnetometry. All measurements were
conducted in zero-field cooled conditions along the $\bigl[1\bar{1}0\bigr]$
in-plane magnetic easy axis,\citep{Bihler_PRB07} and the diamagnetic
background was removed by linear fitting of variable-field magnetic
moment $m\left(H\right)$ data up to $H=\pm50$~$\mathrm{kOe}$ at
$T=5$~$\mathrm{K}$. They revealed an average saturation moment
per $\mathrm{Mn_{Ga}}$ of $m_{\mathrm{sub}}^{\mathrm{sat}}=3.7\pm0.4$~$\mathrm{\mu_{B}}$
in agreement with previous values.\citep{Scarpulla_PRL05} Temperature-dependent
magnetic moment $m\left(T\right)$ data at $H=10$~$\mathrm{Oe}$
revealed $T_{\mathrm{C}}=50\pm1.5$~$\mathrm{K}$, which is well
in line with both previous experimental \citep{Farshchi_SSC06} and
theoretical \citep{Katayama-Yoshida_PSSA07} results. Variable-temperature
sheet resistance $\rho_{\mathrm{s}}\left(T\right)$ measurements showed
similar agreement between samples.

To confirm their structural integrity, samples were characterized
after various irradiation doses. Using IBA, we found that the sheet
concentration of $\mathrm{Mn_{Ga}}$, $c_{\mathrm{s}}$, remains constant
within experimental errors and by SIMS that the Mn distribution is
unaffected. High-resolution transmission electron microscopy and atomic
force microscopy similarly show no qualitative changes with ion irradiation.
Notably, even the sample with the highest irradiation dose shows no
traces of secondary phases.

In order to track the degree of compensation, control samples were
processed in parallel by implanting $\mathrm{Zn^{+}}$ -- a hydrogenic
acceptor in GaP -- to a dose of $1\times10^{16}$~$\mathrm{cm^{-2}}$.
On these, direct measurement of the hole concentration as a function
of irradiation dose is possible using the Hall effect. From this data
we have determined a hole removal rate of $1.1\pm0.1\times10^{3}$
holes per $\mathrm{Ar^{+}}$, or $2.2\pm0.2$ holes per vacancy when
taking into account the simulations. Using this information, we calculate
the relative sheet hole concentration $\Delta p_{\mathrm{s}}$, defined
as the difference in the sheet hole concentration $p_{\mathrm{s}}$
between the unirradiated reference and the irradiated sample.

\begin{figure}
\includegraphics{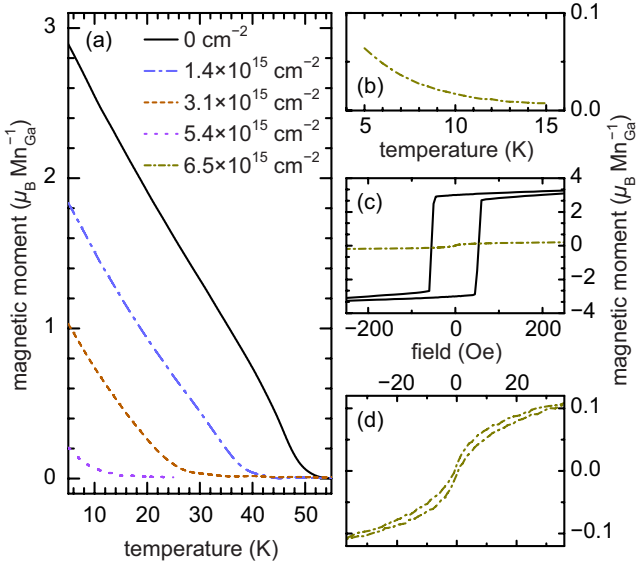}

\caption{\label{fig:MvsT}(color online). (a) and (b) $m\left(T\right)$ at
$H=10$~$\mathrm{Oe}$ for various relative sheet hole concentrations
$\Delta p_{\mathrm{s}}$. (c) $m\left(H\right)$ at $T=5$~$\mathrm{K}$
for as-fabricated and highest-dose irradiated films. (d) Magnetic
hysteresis of the film at an irradiation dose of $5.77\times10^{12}$~$\mathrm{cm^{-2}}$
(also shown in (c)).}

\end{figure}

\begin{figure}
\includegraphics{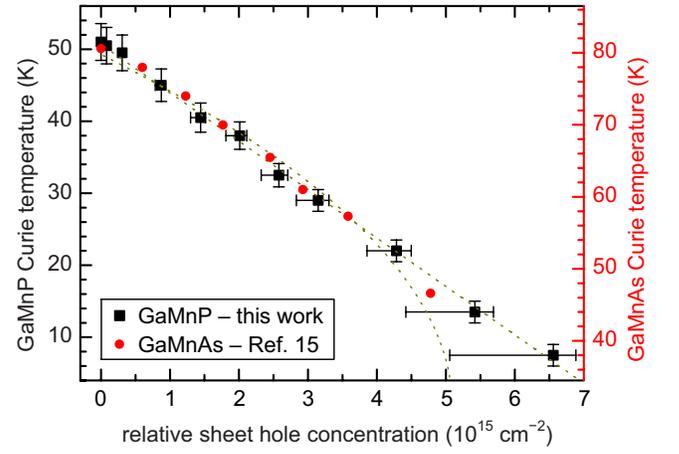}

\caption{\label{fig:Curie}(color online). $T_{\mathrm{C}}$ as a function
of $\Delta p_{\mathrm{s}}$ for Ga$_{1-x}$Mn$_{x}$P samples (squares,
left scale) and Ga$_{0.955}$Mn$_{0.045}$As samples (Ref. \onlinecite{Mayer_PRB10};
circles, right scale). The asymmetric error bars for $\Delta p_{\mathrm{s}}$
reflect saturation effects of vacancy doping. The dotted lines are
the simulated, limiting trends $T_{\mathrm{C}}\propto p$ and $T_{\mathrm{C}}\propto p^{0.5}$.}

\end{figure}

In \prettyref{fig:MvsT}(a-b) we show $m\left(T\right)$ for various
$\Delta p_{\mathrm{s}}$, revealing a monotonic decrease of $T_{\mathrm{C}}$
with $\Delta p_{\mathrm{s}}$. Similarly, we observe a decrease of
$m_{\mathrm{sub}}^{\mathrm{sat}}$ with dose as evidenced in \prettyref{fig:MvsT}(c),
consistent with previous studies of donor- or vacancy doping.\citep{Stone_PRB08,Mayer_PRB10}
We point out that this is in contrast to $m_{\mathrm{sub}}^{\mathrm{sat}}$
being unaffected by hydrogenation,\citep{Bihler_JAP08,Goennenwein_PRL04}
indicating different mechanisms being involved in passivation \textit{versus}
compensation. The dependence of $T_{\mathrm{C}}$ on $\Delta p_{\mathrm{s}}$
is presented in \prettyref{fig:Curie}, revealing a virtually linear
decline with decreasing hole concentration. We note that the highest
irradiation dose of $5.77\times10^{12}$~$\mathrm{cm^{-2}}$ should
be sufficient to fully compensate the Mn acceptors, present at $c_{\mathrm{s}}=5.4\pm0.3\times10^{15}$~$\mathrm{cm^{-2}}$.
However, as apparent from \prettyref{fig:MvsT}(a-d), the films are
FM at all irradiation doses, implying that they remain \textit{p}-type
even for the highest doses. This apparent discrepancy is explained
by the amphoteric defect model (ADM),\citep{Walukiewicz_APL89,Walukiewicz_PRB88}
wherein the defect formation energy strongly depends on the Fermi
level $E_{\mathrm{F}}$, resulting in a saturation of the defect doping-induced
shift in $E_{\mathrm{F}}$ at a material-dependent stabilization level
$E_{\mathrm{FS}}$. This effect becomes dominant in our system for
$\bigl|\Delta p_{\mathrm{s}}\bigr|\gtrsim0.8\, c_{\mathrm{s}}$, considerations
that are reflected in the error bars where appropriate. Furthermore,
the persistence of FM even at these high levels of compensation demonstrates
again that the compensation level of as-fabricated films must be very
low.\citep{Stone_10}

Accounting for the ADM-related compensation effects, we observe the
relation $T_{\mathrm{C}}\propto p^{\gamma}$ with $1>\gamma>0.5$
for Ga$_{1-x}$Mn$_{x}$P. Remarkably, such dependence of $T_{\mathrm{C}}$
on $\Delta p_{\mathrm{s}}$ is nearly identical to that observed in
Ga$_{0.955}$Mn$_{0.045}$As \citep{Mayer_PRB10} films grown by low-temperature
molecular beam epitaxy -- that is, the trend is identical, barring
a certain offset, reminiscent of the similarity in $T_{\mathrm{C}}\left(x\right)$.\citep{Stone_PRL08}
While our $\gamma$ is in a similar range as a \textit{p-d} Zener
model prediction for Ga$_{1-x}$Mn$_{x}$As of $\gamma=0.6\text{--}0.8$,\citep{Nishitani_PRB10,Dietl_PRB01}
the model assumption of uniformly distributed delocalized or weakly
localized holes does not apply to the Ga$_{1-x}$Mn$_{x}$P films
in this study.

\begin{figure}
\includegraphics{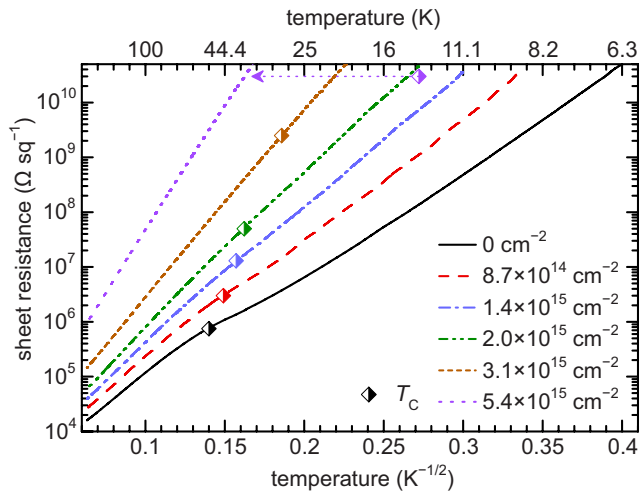}

\caption{\label{fig:res}(color online). $\rho_{\mathrm{s}}$ of Ga$_{1-x}$Mn$_{x}$P
\textit{versus} $T^{-1/2}$ for various relative sheet hole concentrations
$\Delta p_{\mathrm{s}}$. $T_{\mathrm{C}}$ is indicated for each
dose by a diamond, for the highest $\Delta p_{\mathrm{s}}$ supplemented
by an arrow.}

\end{figure}

$\rho_{\mathrm{s}}\left(T\right)$ for Ga$_{1-x}$Mn$_{x}$P samples
with varying levels of compensation is displayed in \prettyref{fig:res}.
Films become orders of magnitude more resistive with increasing irradiation
dose.

The generally applied, phenomenological model in Ga$_{1-x}$Mn$_{x}$P
has been $\rho=\left(\sigma_{\mathrm{free}}\,\exp\left(-\varepsilon_{1}/k_{\mathrm{B}}T\right)\right.+\left.\sigma_{\mathrm{hop}}\,\exp\left(-\varepsilon_{3}/k_{\mathrm{B}}T\right)\right)^{-1}$.\citep{Scarpulla_PRL05}
Here the first term is attributed to thermally activated hole transport
\textit{via} the valence band and the second term to hopping conduction,
previously assumed to take place between nearest neighbors.\citep{Scarpulla_JAP08a,Kaminski_PRB03,Shklovskii_84}
This model reproduces the behavior of samples with varying $x$ which
have not been intentionally compensated.\citep{Farshchi_SSC06} For
the current case of compensated films, however, we find overall better
agreement with activated transport of the form $\rho\propto\exp\left(\varepsilon T^{\lambda}\right)$
with a temperature exponent of $\lambda\sim-0.5$, separated into
a high- and a low-temperature regime characterized by different activation
energies $\varepsilon$. We attribute the general behavior to hopping
conduction, specifically VRH.\citep{Shklovskii_84} That this mechanism
should dominate even at high $T$ for large $\Delta p_{\mathrm{s}}$
is reasonable as the energetic difference between delocalized states
and $E_{\mathrm{F}}$ -- here on the order of the Mn acceptor level
of $0.4$~$\mathrm{eV}$ \citep{Clerjaud_JPC85} -- can easily be
an order of magnitude larger than $k_{\mathrm{B}}T$. At $\Delta p_{\mathrm{s}}\lesssim10^{15}$~$\mathrm{cm^{-2}}$,
VRH is insufficient to describe fully the transport at high $T$.
In this regime, the conduction by holes excited thermally to delocalized
states, as described previously,\citep{Scarpulla_PRL05} contributes
or even dominates. This behavior is qualitatively similar to that
observed in insulating, low-doped Ga$_{1-x}$Mn$_{x}$As \citep{Sheu_PRL07}
and even more so to that in insulating, Sn-codoped Ga$_{1-x}$Mn$_{x}$As.\citep{Satoh_PE01} 

In conclusion, the orders-of-magnitude changes in conductivity and
the much more subtle changes in the magnetic response upon compensation
using $\mathrm{Ar^{+}}$-induced native defects demonstrate the stability
of the hole-mediated FM phase in Ga$_{1-x}$Mn$_{x}$P. While the
electrical behavior of Ga$_{1-x}$Mn$_{x}$P and Ga$_{1-x}$Mn$_{x}$As
at comparable $x$ is dramatically different, these materials display
a remarkably similar $T_{\mathrm{C}}$ dependence on both hole concentration
and Mn content. This indicates similar mechanisms for inter-Mn exchange
in the two systems and places carrier-mediated FM on a continuum of
carrier localization in III-Mn-V DMSs.
\begin{acknowledgments}
The work at Berkeley (sample synthesis, electrical and magnetic characterization,
ion beam analysis) was supported by the Director, Office of Science,
Office of Basic Energy Sciences, Division of Materials Sciences and
Engineering, of the U.S. Department of Energy under Contract No. DE-AC02-05CH11231.
The work at Linz (structural characterization) was supported by the
European Research Council through the FunDMS Advanced Grant within
the “Ideas” 7th Framework Programme of the EC, and by the Austrian
Fonds zur Förderung der wissenschaftlichen Forschung -- FWF (Grants
P22477, P20065, and N107-NAN). We thank R. Jakieła for SIMS measurements.
P.R. S. acknowledges support from an NSF fellowship, T.E. W. from
a Marshall Plan Scholarship.
\end{acknowledgments}

\end{document}